\documentclass[aps,floatfix]{revtex4}
\usepackage{graphics}
\usepackage{epsfig}
\usepackage{subfigure}
\usepackage{amsmath,amsfonts,amssymb,graphicx,times}

\begin{document}

\title{Traffic dynamics in scale-free networks with limited packet-delivering capacity}

\author{Han-Xin Yang$^1$}\email{hxyang@mail.ustc.edu.cn}
\author{Wen-Xu Wang$^2$}
\author{Zhi-Xi Wu$^3$}
\author{Bing-Hong Wang$^1$}\email{bhwang@ustc.edu.cn}

\affiliation{$^{1}$Department of Modern Physics, University of
Science and Technology of China, Hefei 230026,
China\\$^{2}$Department of Electronic Engineering, Arizona State
University, Tempe, Arizona 85287-5706, USA
\\$^{3}$Department of Electronic Engineering, City University of Hong Kong,
Kowloon, Hong Kong, People's Republic of China}

\begin{abstract}
We propose a limited packet-delivering capacity model for traffic
dynamics in scale-free networks. In this model, the total node's
packet-delivering capacity is fixed, and the allocation of
packet-delivering capacity on node $i$ is proportional to
$k_{i}^{\phi}$, where $k_{i}$ is the degree of node $i$ and $\phi$
is a adjustable parameter. We have applied this model on the
shortest path routing strategy as well as the local routing
strategy, and found that there exists an optimal value of parameter
$\phi$ leading to the maximal network capacity under both routing
strategies. We provide some explanations for the emergence of
optimal $\phi$.
\end{abstract}

\maketitle

PACS: 89.75.Hc, 89.20.Hh, 05.70.Fh

Key words: Scale-free networks, Information traffic, Routing
strategy

\section{Introduction}
Since the discovery of small-world effect by Watts and Strogatz
\cite{small-world phenomenon} and scale-free property by
Barab$\acute{a}$si and Albert \cite{scale-free property}, the
structure and dynamics of complex networks have attracted growing
interest and attention from the physics community
\cite{network1,network2,network3,network4,network5}. Due to the
increasing importance of large communication networks such as the
Internet and WWW, information traffic on complex networks has drawn
more and more attention \cite{order
parameter,traffic1,traffic2,traffic3,traffic4,traffic5,traffic6,traffic7,traffic8,traffic9,traffic10,traffic11,traffic12,traffic13,traffic14,traffic15,traffic16,traffic17,traffic18,traffic19,traffic20,traffic21,traffic22,traffic23,traffic24}.
The ultimate goal of studying these large communication networks is
to control the traffic congestion and improve the efficiency of
information transportation.

Researchers have proposed some models to mimic the traffic on
complex networks by introducing packets generating rate $R$ as well
as randomly selected sources and destinations of each packet
\cite{traffic1,traffic2,traffic3,traffic4,traffic5}. In these
models, the capacity of networks is measured by a critical
generating rate $R_{c}$. At this critical rate, a continuous phase
transition from free flow state to congested state occurs. In the
free-flow state, the numbers of created and delivered packets are
balanced, leading to a steady state. While in the jammed state, the
number of accumulated packets increases with time due to the limited
delivering capacity or finite queue length of each node. It has been
found that both network structure and packet routing strategy can
influence the capacity and efficiency of information transportation.

The node packet-delivering capacity, that is, the number of packets
a node can forward to other nodes in each time step, is assumed to
be a constant or proportional to node's degree in most of previous
works. Obviously more packet-delivering capacity can help to
alleviate traffic congestion, but the extending of packet-delivering
capacity will bring economic and technique pressure. So the question
arises: how to rationally allocate the limited packet-delivering
capacity onto nodes in order to maximize the network¡¯s capacity? In
the following, we will explore this question in scale-free networks.

The paper is organized as follows: In Section 2, the traffic model
is introduced. The simulation results are presented and discussed in
Section 3. The conclusion is given in Section 4.

\section{the model}
Recent studies indicate that many communication networks such as the
Internet and WWW are heterogeneous with degree distribution
following the power-law distribution $P(k)\sim k^{-\gamma}$. In this
paper, we use the well-known Barab$\acute{a}$si-Albert (BA)
scale-free network model \cite{scale-free property} as the physical
infrastructure to study information traffic flow. The BA model can
be constructed as follows: starting from $m_{0}$ fully connected
nodes, a new node with $m$ edges is added to the existing graph at
each time step according to preferential attachment, i.e., the
probability $\Pi_{i}$ of being connected to the existing node $i$ is
proportional to the degree $k_{i}$.

Once the network is generated, it remains fixed, and the traffic
dynamics is modeled on top of it as follows: at each time step,
there are $R$ packets generated in the system, with randomly chosen
sources and destinations. All the nodes act as both hosts and
routers and node $i$ can deliver at most $C_{i}$ packets per time
step towards their destinations. Once a packet arrives at its
destination, it will be removed from the system. The queue length of
each node is assumed to be unlimited and the FIFO (first in first
out) discipline is applied at each queue \cite{traffic1,traffic2}.

Packets can be delivered according to different routing strategies.
In this paper, we considers the network traffic in the cases of both
the shortest path and local routing strategy. The local routing
strategy \cite{traffic9} can be described as follows. Each node
performs a local search among its neighbors. If the packet's
destination is found within the searched area, i.e., among the
node's immediate neighbors, it is delivered directly to its target.
Otherwise, it is forwarded to a neighbor node $i$, according to the
probability:
\begin{equation}
\Pi_{i}=\frac{k_{i}^{\alpha}}{\sum_{j} k_{j}^{\alpha}},
\end{equation}
where the sum runs over the neighbors (searched area) of the
searching node, $k_{i}$ is the degree of node $i$ and $\alpha$ is an
adjustable parameter. The average packet-delivering capacity of the
network is:
\begin{equation}
\langle C \rangle=\frac{1}{N}\sum_{i=1}^{N}C_{i}.
\end{equation}
Based on economic and technique considerations, it's significative
to investigate how to allocate packet-delivering capacity onto nodes
when $\langle C \rangle$ is fixed. Since scale-free network is
heterogeneous, packet-delivering capacity can be allocated in the
form of:
\begin{equation}
C_{i}=N\langle C
\rangle\frac{k_{i}^{\phi}}{\sum_{j=1}^{N}k_{j}^{\phi}},
\end{equation}
where $\phi$ is an adjustable parameter. For $\phi>0$ ($\phi<0$),
nodes with higher (smaller) degrees have larger packet-delivering
capacity. When $\phi=0$, all nodes have the same packet-delivering
capacity. Noting that $C_{i}$ may be an integer plus a fractional
part, the fractional part is implemented as the probability of
delivering additional packets in a time step.

\section{Simulation results}
In order to characterize the network capacity, we use the order
parameter presented in Ref. \cite{order parameter}:
\begin{equation}
\eta(R)=\lim_{t\rightarrow\infty}\frac{1}{R}\frac{\langle\Delta
N_{p}\rangle}{\Delta t},
\end{equation}
where $\Delta N_{p}=N_{p}(t+\Delta t)-N_{p}(t)$, $\langle \cdot
\cdot \cdot \rangle$ indicates the average over time windows of
width $\Delta t$, and $N_{p}(t)$ represents the number of data
packets within the network at time $t$. With increasing packet
generation rate $R$, there will be a critical value of $R_{c}$ that
characterizes the traffic phase transition from free flow to a
congested state. When $R < R_{c}$, $\langle \Delta N_{p}\rangle= 0$
and $\eta(R)= 0$, corresponding to the case of free-flow state.
However, for $R>R_{c}$, $\eta(R)$ is a constant larger than zero,
the packets will continuously pile up within the network and the
system will collapse ultimately. Therefore $R_{c}$ is the maximal
generating rate under which the system can maintain its normal and
efficient functioning. Thus the overall capacity of the system can
be measured by $R_{c}$.

Figure 1 reports the order parameter $\eta$ versus generating rate
$R$ for different parameter $\phi$ under the shortest path root
strategy. One can see that, for all different $\phi$, $\eta$ is
approximately zero when $R$ is small; it suddenly increases when $R$
is larger than the critical point $R_{c}$. It is clear to find that
the capacity of the system is not the same for different $\phi$.

Figure 2 shows $R_{c}$ versus $\phi$ for different $\langle C
\rangle$ under the shortest path routing strategy. Interestingly, we
find $R_{c}$ is not a monotonic function of $\phi$. There exists an
optimal value of $\phi$ (positive) corresponding to the largest
$R_{c}$, which means neither the uniform allocation nor the
extremely uneven distribution can maximize network capacity. The
emergence of optimal $\phi$ can be explained by betweenness
centrality (BC) distributions in scale-free network
\cite{BC1,BC2,BC3}. The BC of a node $v$ is defined as:
\begin{equation}
g(v)=\sum_{s\neq t}\frac{\sigma_{st}(v)}{\sigma_{st}},
\end{equation}
where $\sigma_{st}$ is the number of shortest paths going from $s$
to $t$ and $\sigma_{st}(v)$ is the number of shortest paths going
from $s$ to $t$ and passing through $v$. BC gives an estimate of the
traffic load on nodes when packets are forwarded following their
shortest paths. For scale-free networks it has been shown that
relationship between betweenness centrality and degree obeys
power-law form: $g(k)\sim k^{\mu}$, and large-degree nodes endure
much heavier traffic load than that of small-degree nodes. Figure 3
shows that the exponent $\mu=1.33$ when the network parameters are
$m_{0}=m=3$, $N=1000$. Interestingly, we find this value of exponent
is approximately equal to the optimal $\phi_{opt}=1.3$ observed in
Fig. 2.

\begin{figure}
\scalebox{0.8}[0.7]{\includegraphics{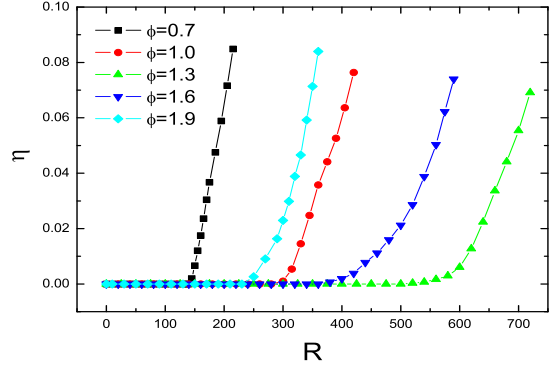}}
\caption{\label{fig:epsart} The order parameter $\eta$ versus $R$
for different $\phi$ under the shortest path root strategy. Average
packet-delivering capacity of the network is $\langle C \rangle=3$
and the network parameters are $N=1000$, $m_{0}=m=3$.  }
\end{figure}

\begin{figure}
\scalebox{0.8}[0.75]{\includegraphics{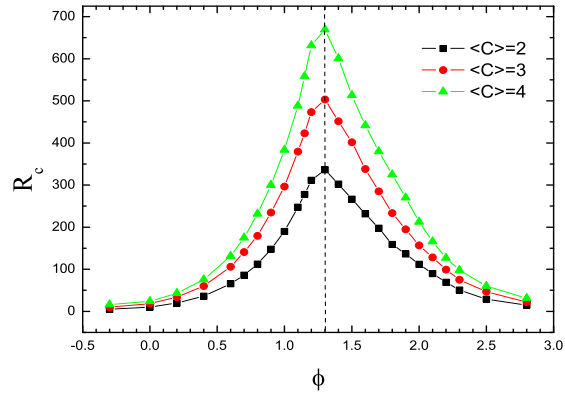}}
\caption{\label{fig:epsart} The critical $R_{c}$ versus $\phi$ for
different $\langle C \rangle$ under the shortest path routing
strategy. The network parameters are $N=1000$, $m_{0}=m=3$.}
\end{figure}

\begin{figure}
\scalebox{0.8}[0.7]{\includegraphics{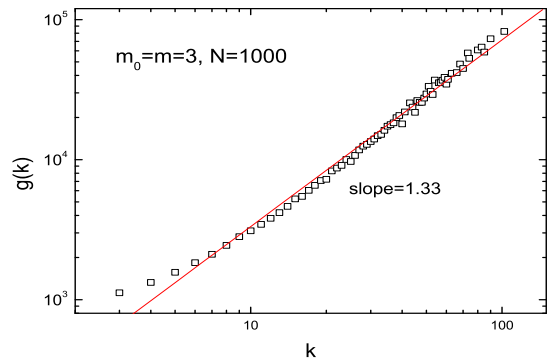}}
\caption{\label{fig:epsart} Log-Log plot of betweenness centrality
$g(k)$ versus degree $k$. The network parameters are $N=1000$,
$m_{0}=m=3$. The fitted line has a slope $\mu=1.33$.}
\end{figure}

To understand why $\phi_{opt} = \mu$ results in the maximal network
capacity under the shortest path routing strategy, we investigate
the queue length of a node $n(k)$ as a function of its degree $k$ in
the congested state ($R>R_{c}$). The queue length of a node is
defined as the number of packets in the queue of that node. For
$\phi$ is small, i.e., $\phi=0$, large-degree nodes do not get
enough packet-delivering capacity while small-degree nodes have
redundant packet-delivering capacity which exceeds their actual load
. As shown in Fig. 4(a), the queue length of large-degree nodes
becomes longer and longer while at the same time small-degree nodes
almost have no packets on their queue. Contrarily, if $\phi$ is very
large, i.e., $\phi=2.5$, most of packet-delivering capacity is
allocated to a few large-degree nodes and many small-degree nodes
have too little packet-delivering capacity to fully dispose the load
on them. As a result, packets continuously pile up on small-degree
nodes (see Fig. 4(b)). In order to make full use of limited
packet-delivering capacity and avoid congestion on a few nodes, the
load distribution should be consistent with the packet-delivering
capacity distribution, that is, $\phi_{opt} = \mu$. This average
effect results in the maximal network capacity.

\begin{figure}
\scalebox{0.9}[0.8]{\includegraphics{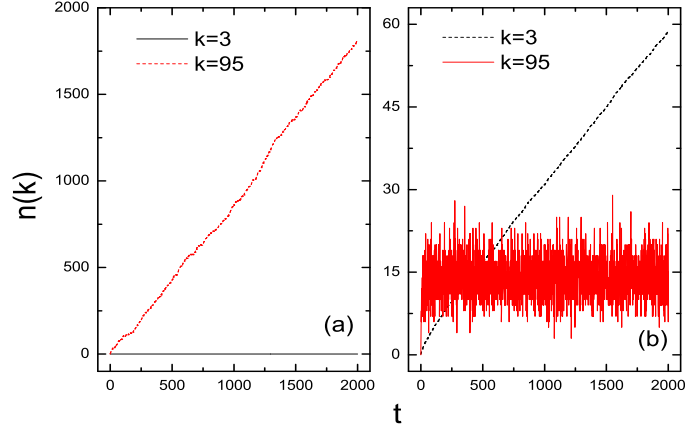}}
\caption{\label{fig:epsart} Evolution of queue length $n(k)$ for
different degree $k$ under the shortest path routing strategy.
$N=1000$, $m_{0}=m=3$, $\langle C \rangle=3$. The smallest degree
$k=3$ and the largest degree $k=95$ in the BA network. (a)
$R=25>R_{c}=18$ for $\phi=0.0$ and (b) $R=100>R_{c}=47$ for
$\phi=2.5$.}
\end{figure}

\begin{figure}
\scalebox{0.8}[0.8]{\includegraphics{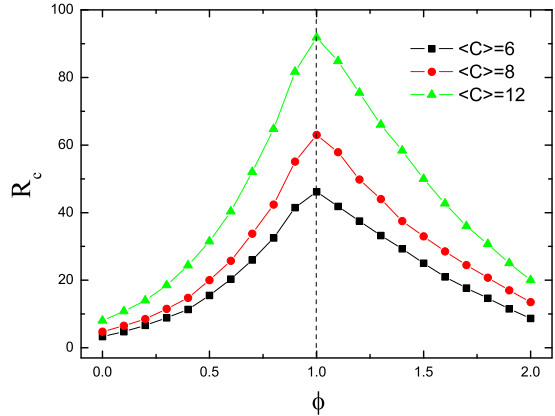}}
\caption{\label{fig:epsart} The critical $R_{c}$ versus $\phi$ for
different $\langle C \rangle$ under the local routing strategy
($\alpha=0$). The network parameters are $N=1000$, $m_{0}=m=4$. }
\end{figure}

Next we investigate the behavior of $R_{c}$ versus $\phi$ for
different $\langle C \rangle$ under the local routing strategy. As
shown in Fig. 5, there also exists an optimal value of $\phi$
corresponding to the largest $R_{c}$. For $\alpha=0$, $\phi_{opt}=
1$. The optimal value of $\phi$ corresponding to different $\alpha$
is shown in Fig. 6(a). It's found that $\phi_{opt} \approx 1+\alpha$
for the local routing strategy. According to the analysis in Ref.
\cite{traffic9}, the relationship between the queue length and
degree is a power-law form: $n(k)\sim k^{1+\alpha}$ in the free flow
state. To maximize the network capacity, the relationship between
the packet-delivering capacity and degree also obeys the same
power-law form. Furthermore, we study the maximum $R_{c}$ as a
function of $\alpha$ (Fig. 6(b)). One can find that there also
exists nonmonotonous behavior with a peak at about $a=0.2$.

\begin{figure}
\scalebox{0.9}[0.8]{\includegraphics{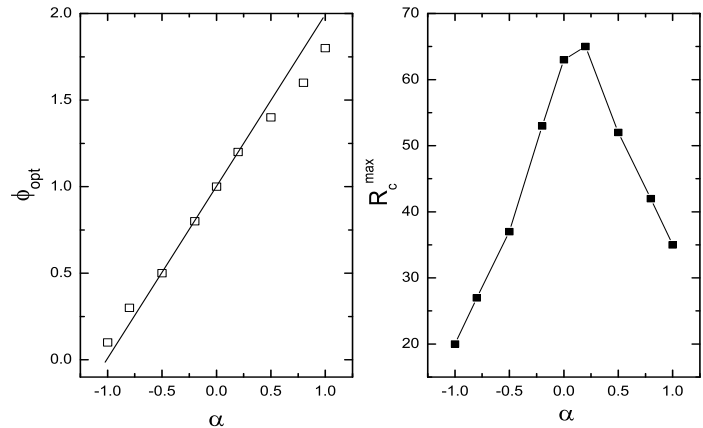}}
\caption{\label{fig:epsart} (a) The optimal value of $\phi$ for
different $\alpha$ under the local routing strategy. The line is the
the theoretical prediction. (b) The maximum $R_{c}$ as a function of
$\alpha$. The network parameters are $N=1000$, $m_{0}=m=4$. The
average packet-delivering capacity of the network $\langle C
\rangle=8$.}
\end{figure}

\section{conclusion}

In conclusion, we have investigated how to rationally allocate
packet-delivering capacity onto nodes in the BA scale-free network
when the sum of all nodes' packet-delivering capacity is fixed. A
tunable parameter is introduced, governing node's packet-delivering
capacity based on its degree. Interestingly, we find there exists an
optimal value of parameter $\phi$ leading to the maximal network
capacity. We provide some explanations for the emergence of optimal
$\phi$ by investigating betweenness centrality distribution in the
shortest path routing strategy and the queue length distribution in
the local routing strategy. Our work may be helpful for designing
realistic communication network.

\section*{ACKNOWLEDGMENTS}
This work is funded by the National Basic Research Program of China
(973 Program No.2006CB705500), the National Natural Science
Foundation of China (Grant Nos. 60744003, 10635040, 10532060,
10472116), by the Special Research Funds for Theoretical Physics
Frontier Problems (NSFC No.10547004 and A0524701), by the President
Funding of Chinese Academy of Science, and by the Specialized
Research Fund for the Doctoral Program of Higher Education of China.


\begin{thebibliography}{ref1}

\bibitem{small-world phenomenon} D. J. Watts, S. H. Strogatz, Nature 393 (1998) 440.
\bibitem{scale-free property} A. L. Barab$\acute{a}$si, R. Albert, Science 286 (1999) 509.


\bibitem{network1} R. Albert, A. L. Barab$\acute{a}$si, Rev. Modern. Phys. 74 (2002) 47.
\bibitem{network2} M. E. J. Newman, Phys. Rev. E 64 (2001) 016132.
\bibitem{network3} M. E. J. Newman, SIAM Rev. 45 (2003) 167.
\bibitem{network4} R. Pastor-Satorras, A. Vespignani, Evolution and Structure of the Internet: A Statistical Physics Approach, Cambridge University Press, Cambridge,
2004.
\bibitem{network5} S. Boccaletti, V. Latora, Y. Moreno, M. Chavez, D. U.
Hwang, Phys. Rep. 424 (2006) 175.

\bibitem{order parameter} A. Arenas, A. D\'{\i}z-Guilera, R. Guimer$\grave{a}$, Phys.
Rev. Lett. 86 (2001) 3196.
\bibitem{traffic1} B. Tadi$\acute{c}$, S. Thurner, G. J. Rodgers, Phys. Rev. E 69
(2004) 036102.
\bibitem{traffic2} L. Zhao, Y. C. Lai, K. Park, N. Ye, Phys. Rev. E
71 (2005) 026125.
\bibitem{traffic3} G. Mukherjee and S. S. Manna, Phys. Rev. E 71 (2005) 066108.
\bibitem{traffic4} R. Guimer$\grave{a}$, A. D\'{\i}az-Guilera, F. Vega-Redondo, A. Cabrales,
A. Arenas, Phys. Rev. Lett. 89 (2002) 248701.
\bibitem{traffic5} R. Guimer$\grave{a}$, A. Arenas, A. D\'{\i}az-Guilera, F. Giralt, Phys.
Rev. E 66 (2002) 026704.


\bibitem{traffic6} P. Echenique, J. G$\acute{o}$mez-Garde$\tilde{n}$es, Y. Moreno, Phys. Rev. E 70 (2004)
056105.
\bibitem{traffic7} D. J. Ashton, T. C. Jarrett, N. F. Johnson, Phys. Rev. Lett. 94 (2005)
058701.
\bibitem{traffic8} G. Yan, T. Zhou, B. Hu, Z. Q. Fu, B. H. Wang, Phys. Rev. E 73 (2006) 046108.
\bibitem{traffic9} W. X. Wang, B. H. Wang, C. Y. Yin, Y. B. Xie, T. Zhou, Phys. Rev. E. 73 (2006) 026111.
\bibitem{traffic10} W. X. Wang, C. Y. Yin, G. Yan, B. H. Wang, Phys. Rev. E. 74 (2006)
016101.
\bibitem{traffic11} C. Y. Yin, B. H. Wang, W. X. Wang, G. Yan, H. J. Yang, Eur. Phys. J. B 49 (2006) 205.
\bibitem{traffic12} C. Y. Yin, B. H. Wang, W. X. Wang, T. Zhou, H. J. Yang, Phys. Lett. A 351 (2006) 220.
\bibitem{traffic13} M. B. Hu, W. X. Wang, R. Jiang, Q. S. Wu, Y. H. Wu, Europhys. Lett. 79 (2007)
14003.
\bibitem{traffic14} M. B. Hu, W. X. Wang, R. Jiang, Q. S. Wu, Y. H. Wu, Phys. Rev. E 75 (2007) 036102.
\bibitem{traffic15} Z. Liu, M. B. Hu, R. Jiang, W. X. Wang, Q. S. Wu, Phys. Rev. E 76 (2007)
037101.
\bibitem{traffic16} Z. X. Wu, W. X. Wang, K. H. Yeung, New Journal of Physics 10 (2008)
023025.
\bibitem{traffic17} H. Zhang, Z. H. Liu, M. Tang, P. M. Hui, Phys. Lett. A 364 (2007)
177.
\bibitem{traffic18} Z. H. Liu, W. C. Ma, H. A. Zhang, Y. Sun, P. M. Hui, Physica A 370 (2006) 843.
\bibitem{traffic19} X. Ling, R. Jiang, X. Wang, M. B. Hu, Q. S. Wu, Physica A 387
(2008) 4709.
\bibitem{traffic20} M. B. Hu, R. Jiang, Y. H. Wu, W. X. Wang, Q. S.
Wu, Physica A 387 (2008) 4967.
\bibitem{traffic21} S. Valverde, R. V. Sol$\acute{e}$, Eur. Phys. J. B 38 (2004) 245.
\bibitem{traffic22} V. Cholvi, V. Laderas, L. L$\acute{o}$pez, A. Fern$\acute{a}$ndez, Phys. Rev.
E 71 (2005) 035103(R).
\bibitem{traffic23} B. K. Singh, N. Gupte, Phys. Rev. E 71 (2005)
055103(R).
\bibitem{traffic24} X. Gong, L. Kun, C. H. Lai, Europhys. Lett. 83 (2008)
28001.


\bibitem{BC1} M. E. J. Newman, M. Girvan, Phys. Rev. E 69 (2004) 026113.
\bibitem{BC2} K. I. Goh, B. Kahng, D. Kim, Phys. Rev. Lett. 87
(2001) 278701.
\bibitem{BC3} M. Barth$\acute{e}$lemy, Eur. Phys. J. B 38 (2003) 163.


\end{thebibliography}
\end{document}